\documentclass[conference]{IEEEtran}
\usepackage{amsmath,amssymb,mathtools,amsthm,bm}
\usepackage{physics}
\usepackage{bbm}
\usepackage{cite}
\usepackage{dsfont}
\usepackage{enumitem}
\usepackage{booktabs}
\usepackage[caption=false,font=footnotesize]{subfig}
\usepackage[hidelinks]{hyperref}
\usepackage{doi}
\usepackage{algorithm}
\usepackage{algpseudocode}
\usepackage{tikz}
\usetikzlibrary{shapes.geometric, arrows.meta, positioning, calc}
\usepackage{balance}
\usepackage{needspace}

\newcommand{\Cov}{\mathrm{Cov}}

\begin{document}
%
\title{Spatial Load Correlation in AI Data-Center-Dominated Power Systems \vspace{-0.5em}}

\author{\IEEEauthorblockN{Chandan Chaudhary,~\emph{Student Member, IEEE}, Alaaeldein Abdelkader,~\emph{Student Member, IEEE}, \\ Yansong Pei,~\emph{Member, IEEE}, Mohammed Benidris,~\emph{Senior Member, IEEE}, and Joydeep Mitra,~\emph{Fellow, IEEE}}
\IEEEauthorblockA{Electrical and Computer Engineering, Michigan State University, East Lansing, MI 48824, USA}
Emails: chaud152@msu.edu, abdelk15@msu.edu, peiyanso@msu.edu, benidris@msu.edu, and mitraj@msu.edu \vspace{-2em}}


{\maketitle
\vspace{-1.5em}
}
\begin{abstract}
The proliferation of large-scale data centers introduces spatially correlated demand profiles that challenge the long-standing assumption of statistical independence of loads in power system analysis. This paper examines the emergence of such load correlations and evaluates their impact on data-center-dominated grids. Analytical derivations reveal that correlated load fluctuations amplify aggregate stochastic disturbances, reduce voltage stability margins through weakened reactive power stiffness, and degrade frequency stability margin by erosion of natural load diversity effects. Real-time digital simulation studies confirm that moderate spatial correlation in distributed data centers produces simultaneous frequency deviations and voltage fluctuations across multiple buses. The findings offer transmission system operators a physics-based perspective to interpret emerging oscillatory phenomena and establish stability planning criteria grounded in measurable load-correlation structures rather than traditional diversity assumptions.
\end{abstract}

%

\IEEEpeerreviewmaketitle
\vspace{-0.55em}
\section{Introduction}
The rapid expansion of AI and high-performance computing (HPC) has fundamentally reshaped electricity demand across major interconnections. Data centers have emerged as one of the most consequential load classes directly interfacing with bulk power system. Recent projections reveal that global data centers could demand up to 945~TWh annually by 2030, with AI-driven workloads accounting for up to 12\% of U.S. electricity consumption by 2028~\cite{shehabi2024usdc, nerc2025largeloads}. Modern hyperscale facilities operate at extreme power densities, often exceeding 100~MW per site \cite{nerc2025largeloads, QuintEtAl2025LargeLoad}, and some campuses are now planned at the gigawatt scale. Such facilities exhibit sharp power fluctuations arising from synchronized compute and communication phases in large-scale training jobs~\cite{go2025characterizing, QuintEtAl2025LargeLoad}.

A defining characteristic of this evolution is geographic concentration. Northern Virginia's “Data Center Alley” now exceeds 2.5 GW of active demand, and the addition of new AI-data-center campuses produces highly localized, synchronized load patterns that intensify transmission stress~\cite{nerc2025largeloads, QuintEtAl2025LargeLoad}. This clustering forms demand hubs that impose intense stress on regional transmission infrastructure and invalidate the load diversity assumptions of power system planning and operation.

Data centers interface with the grid primarily through voltage-source converters, which include double-conversion UPS systems, rack-level rectifiers, and high-density server power supplies~\cite{xiong2020modeling,Chaudhary2025DataCenterStability,sun2021development}. This converter-dominated architecture creates loads with ultra-low inertia, weak damping, and negative impedance across multiple frequency bands, which decouple demand from inertial buffering and differ from electromechanical loads. Field measurements reveal low-frequency resonances near 11~Hz and high-frequency oscillations in the 5--10~kHz range at rack, triggered by control delays and DC bus dynamics~\cite{sun2022dataI, sun2022dataII}. Real-world events, such as 14.7--14.8~Hz oscillations in Dominion Energy, show that unplanned excitations from data center operations destabilize hidden dynamic modes in weak grids~\cite{mishra2025understanding}.

AI and HPC load profiles demonstrate that these loads generate megawatt-scale ramps within seconds during synchronized compute-communication cycles, with fluctuation spectra capable of exciting inter-area electromechanical modes when aligned with weakly damped grid resonances~\cite{QuintEtAl2025LargeLoad, Chaudhary2025DataCenterStability, ko2025wide}. Multiple data centers operating under shared orchestration platforms, thermal constraints, or workload scheduling exhibit correlated power variations across distinct network buses. These loads induce voltage-sensitive reductions, trigger frequency excursions, and exacerbate stability risks in converter-rich grids~\cite{nerc2025largeloads, Chaudhary2025DataCenterStability}.

Although existing studies provide insight into single-site data-center dynamics, analytical tools that capture correlation across geographically dispersed facilities and translate those effects to system-level behavior remain scarce. Planning criteria assume load independence and overlook impacts of correlated load variations in stability thresholds or reserve requirements. This paper addresses these gaps in three ways: (1) defining spatial load correlation mathematically and identifying the physical mechanisms that create it; (2) deriving analytical relationships that link load correlation to voltage stability, frequency response, and oscillation amplification; and (3) validating the theoretical framework through real-time digital simulations that show even moderate correlation can produce simultaneous disturbances across multiple buses.

The remainder of this paper is organized as follows. Section~\ref{sec:Correlated_load} introduces the concept of correlated load. The physical mechanisms giving rise to load correlation are detailed in Section~\ref{sec:Physical_mechanism}. Section~\ref{sec:Impacts} quantifies the impacts of correlated load on voltage stability, frequency dynamics, inter-area oscillations, and reserve planning requirements. Section~\ref{sec:RTDS} presents real-time digital simulation studies on the IEEE~39-bus system. Finally, Section~\ref{sec:Conclusion} concludes the paper and provides direction for future research in this sector.

\vspace{-0.5em}
\section{Correlated Load}
\label{sec:Correlated_load}
\vspace{-0.25em}
Power system analysis has relied on the premise that electrical loads at different network locations fluctuate independently. This consideration, valid for dispersed and uncoordinated residential or industrial demand, offers analytical simplicity and justifies the use of diversity factors in planning and stability studies. Each bus-level demand is modeled as an uncorrelated stochastic process $\{P_i(t)\}_{i=1}^N$, which ensures aggregate variance grows linearly with the number of loads. This naturally mitigates system-wide fluctuations.

This independence simplifies forecasting, modal analysis, and infrastructure planning as each load acts as an isolated stochastic source. This allows linearized small-signal models and probabilistic methods to treat buses separately. Mathematically, independence requires that for $i \neq j$,
\begin{equation}
\mathbb{E}[P_i(t)P_j(t')]\! = \! \mathbb{E}[P_i(t)]\mathbb{E}[P_j(t')], \mathrm{Corr}(P_i(t),P_j(t'))\! =\! 0
\label{eq:independence}
\end{equation}
a condition increasingly violated by the rise of synchronized and digitally managed facilities.

\vspace{-0.25em}
\subsection{Spatially Correlated Load Behavior}
Modern grids are comprised of converter-dominated entities such as data centers, EV fleets, and smart campuses, that operate under centralized control or shared environmental stimuli. Their coordinated behavior induces temporal correlation among power injections at geographically distinct buses, which gives rise to \emph{spatially correlated load}. From a system perspective, such correlation emerges only when fluctuations at multiple nodes evolve coherently in time. 

A single, time-varying load at one bus, regardless of the internal synchronization among its subsystems, does not constitute spatial correlation because the grid perceives only one aggregated injection. A facility may be composed of multiple subsystems that operate in synchrony. If the facility connects to the transmission network through a single node $i$, the grid observes only the total power injection as
a single stochastic process. Internal synchronization among subsystems, such as GPU racks or HVAC controls, remains electrically invisible, thus yields $\text{Corr}(P_i(t),P_j(t'))=0$ for all $j \neq i$. 

\begin{figure}[htbp]
\centering
\vspace{-1.25 em}
\begin{tikzpicture}[scale=0.78, every node/.style={scale=0.78}]
\begin{scope}[shift={(-0.3,0)}]
\node[font=\scriptsize, text=red!70!black] at (3.5, 5.8) {Transmission Level};
\node[font=\scriptsize, text=blue!70!black] at (3.5, 3.7) {Subtransmission};
\node[font=\scriptsize, text=blue!70!black] at (3.5, 3.45) {Level};

\node[circle, draw, thick, minimum size=0.8cm, fill=red!12] (B1) at (0,5.0) {\small Bus 1};
\node[circle, draw, thick, minimum size=0.8cm, fill=red!12] (B2) at (3.5,5.0) {\small Bus 2};
\node[circle, draw, thick, minimum size=0.8cm, fill=red!12] (B3) at (7.0,5.0) {\small Bus 3};

\draw[very thick, red!60!black] (B1)--(B2) node[midway,above,font=\small]{230 kV};
\draw[very thick, red!60!black] (B2)--(B3) node[midway,above,font=\small]{230 kV};

\foreach \x in {0, 2.5, 4.5, 7.0}{
  \node[circle,draw,thick,minimum size=0.25cm,fill=white] at (\x,4.0){};
  \draw[thick](\x,3.8)--(\x,4.2);
  \draw[thick](\x-0.18,4.0)--(\x+0.18,4.0);
}
\draw[thick,blue!60!black](B1)--(0,4.0);
\draw[thick,blue!60!black](B2)--(2.5,4.0);
\draw[thick,blue!60!black](B2)--(4.5,4.0);
\draw[thick,blue!60!black](B3)--(7.0,4.0);

\node[circle,draw,thick,fill=blue!18,minimum size=0.8cm] (S1) at (0,2.8) {\small Sub.1};
\node[circle,draw,thick,fill=blue!18,minimum size=0.8cm] (S2) at (2.5,2.8) {\small Sub.2};
\node[circle,draw,thick,fill=blue!18,minimum size=0.8cm] (S3) at (4.5,2.8) {\small Sub.3};
\node[circle,draw,thick,fill=blue!18,minimum size=0.8cm] (S4) at (7.0,2.8) {\small Sub.4};

\draw[thick,blue!60!black](0,4.0)--(S1);
\draw[thick,blue!60!black](2.5,4.0)--(S2);
\draw[thick,blue!60!black](4.5,4.0)--(S3);
\draw[thick,blue!60!black](7.0,4.0)--(S4);
\draw[thick,blue!50](S1)--(S2) node[midway, above,font=\scriptsize]{69–138 kV};
\draw[thick,blue!50](S3)--(S4) node[midway, above,font=\scriptsize]{69–138 kV};

\node[rectangle,draw,thick,fill=orange!20,minimum width=1.9cm,minimum height=1cm,align=center] (DC1) at (0,0.7) {\scriptsize\textbf{Data Center 1}\\[-1pt]\scriptsize150 MW\\[-1pt]\scriptsize AI Training};
\node[rectangle,draw,thick,fill=orange!20,minimum width=1.9cm,minimum height=1cm,align=center] (DC2) at (2.5,0.7) {\scriptsize\textbf{Data Center 2}\\[-1pt]\scriptsize180 MW\\[-1pt]\scriptsize AI Training};
\node[rectangle,draw,thick,fill=green!15,minimum width=1.7cm,minimum height=0.9cm,align=center] (Load) at (4.5,0.7) {\scriptsize Independent\\[-1pt]\scriptsize Load 50 MW};
\node[rectangle,draw,thick,fill=orange!20,minimum width=1.9cm,minimum height=1cm,align=center] (DC3) at (7.0,0.7) {\scriptsize\textbf{Data Center 3}\\[-1pt]\scriptsize120 MW\\[-1pt]\scriptsize AI Inference};

\draw[->,thick,orange!70!black](DC1)--(S1)node[midway,left,xshift=-2pt,font=\scriptsize]{$P_1$};
\draw[->,thick,orange!70!black](DC2)--(S2)node[midway,right,xshift=2pt,font=\scriptsize]{$P_2$};
\draw[->,thick,orange!70!black](DC3)--(S4)node[midway,left,font=\scriptsize]{$P_3$};
\draw[->,thick,green!60!black](Load)--(S3)node[midway,right,font=\scriptsize]{$P_4$};

\coordinate (DC2_bottom_40L) at ($(DC2.south west)!0.40!(DC2.south east)$);
\coordinate (DC2_bottom_60L) at ($(DC2.south west)!0.60!(DC2.south east)$);

\draw[<->,thick,red!65,dashed]
  (DC1.south) to[bend right=20]
  node[midway,below=4pt,font=\scriptsize,text=red!70!black]{Sync'd Workload, Weather}
  (DC2_bottom_40L);

\draw[<->,thick,red!65,dashed]
  (DC2_bottom_60L) to[bend right=20]
  node[midway,below=1.5pt,font=\scriptsize,text=red!70!black]{Correlated}
  (DC3.south);

\node[font=\scriptsize] at (0.9,-0.9) {$\downarrow P_1(t)=\bar{P}+A\cos(\omega_ct)$};
\node[font=\scriptsize] at (3.0,-1.25) {$\downarrow P_2(t)=\bar{P}+A\cos(\omega_ct+\phi_{12})$};
\node[font=\scriptsize] at (7.0,-0.9) {$\downarrow P_3(t)=\bar{P}+A\cos(\omega_ct+\phi_{13})$};

\end{scope}

\begin{scope}[shift={(9,0)}, scale=0.70]

\node[align=center, font=\scriptsize] at (-0.7,5.5) {
$\mathbf{C} \!=\!
\left[
\begin{array}{@{}cccc@{}}
\! 1 & \! C_{12} \!& \! C_{13} \! & \! 0 \\
\!C_{12} \!&\! 1\! & \!C_{23}\! & \!0 \\
\!C_{13} \!& \!C_{23} & 1 & 0 \\
\!0 &\! 0\! & \!0 & \!1
\end{array}
\right]
$};

    \node[draw, thick, fill=red!10,
          minimum width=2cm, minimum height=2.25cm]
          (inset) at (0,1.2) {};

    \node[font=\scriptsize, text=black] at (0,2.4)
        {Internal Sync};

    \node[rectangle, draw, fill=orange!30,
          minimum width=0.9cm, minimum height=0.2cm]
          at (-0.68,1.8) {\tiny $\text{R}_1$};

    \node[rectangle, draw, fill=orange!30,
          minimum width=0.9cm, minimum height=0.2cm]
          at (-0.68,1.15) {\tiny $\text{R}_2$};

    \node at (-0.65,0.75) {\scriptsize$\cdots$};

    \node[rectangle, draw, fill=orange!30,
          minimum width=0.9cm, minimum height=0.2cm]
          at (-0.68,0.35) {\tiny $\text{R}_n$};

    \node[rectangle, draw, fill=cyan!30,
          minimum width=0.7cm, minimum height=0.32cm]
          at (0.75,1.20) {\tiny HVAC};

    \node[font=\tiny, text=red!80!black]
         at (0,-0.15) {No spatial correlation};

    \node[font=\tiny] at (0.55,0.40) {R = Rack};

\end{scope}

\draw[->, thick, blue!60!black]
      (DC3.east) -- ++(0.33,0.1)
      node[midway, above, font=\scriptsize] {};

\end{tikzpicture}
\vspace{-2.2em}
\caption{Representation of multi-site correlated data-center loads with rack-level and HVAC synchronization.}
\label{fig:grid_correlated}
\vspace{-1.5em}
\end{figure}
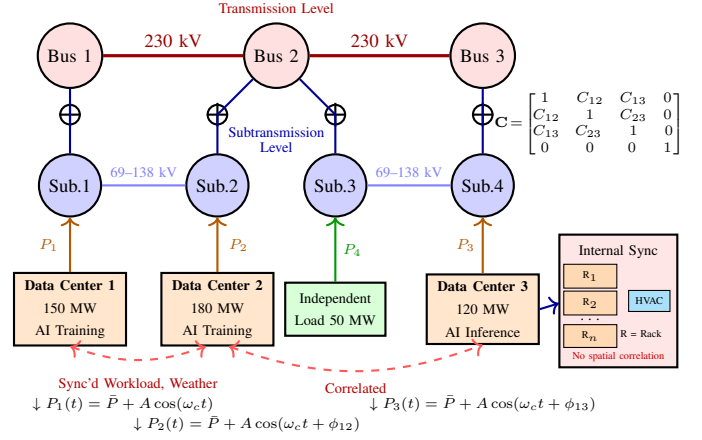

Spatial correlation instead occurs when loads at distinct buses $i$ and $j$ (with $\mathbf{Z}_{ij} \neq 0$, $\mathbf{Z}$ is the network impedance matrix) fluctuate coherently. The correlation strength is captured by the cross-correlation function,
\begin{equation}
\rho_{ij}(\tau) =
\frac{\mathbb{E}[(P_i(t)-\bar{P}_i)(P_j(t+\tau)-\bar{P}_j)]}{\sigma_i \sigma_j},
\label{eq:cross_correlation_reduced}
\end{equation}
and a significant correlation exists when $|\rho_{ij}(\tau)|>\epsilon$ for some $\tau \in [-\tau_{\max},\tau_{\max}]$. The resulting correlation matrix
\vspace{-0.5em}
\begin{equation}
C_{ij} = \max_{\tau \in [-\tau_{\max},\tau_{\max}]} |\rho_{ij}(\tau)|
\label{eq:correlation_matrix_reduced}
\end{equation}
is symmetric with $C_{ii}=1$ and $C_{ij}\in[0,1]$. Strong off-diagonal terms indicate synchronized fluctuations among distinct buses, a phenomenon increasingly observed in data-center-dominated grids. As illustrated in Fig.~\ref{fig:grid_correlated}, synchronized facilities connected to distinct buses can generate strong off-diagonal correlation terms $C_{ij}$, even when neighboring loads remain independent. This spatial structure, characterized by selective non-zero off-diagonal elements in $\mathbf{C}$, fundamentally differentiates correlated load phenomena from single-site temporal variability and necessitates analytical frameworks that extend beyond conventional power flow formulations.

\vspace{-2em}
\section{Physical Mechanisms of Load Correlation}
\label{sec:Physical_mechanism}
Correlated load in modern power systems stems from physical, operational, and environmental couplings among large digitally managed facilities. This section examines the physical mechanisms that induce such correlations across grid buses. In data-center-dominated networks, these mechanisms fall into three categories: operational synchronization, converter-mediated electrical coupling, and environmental forcing.
\vspace{-0.5em}
\subsection{Cross-Facility Workload Orchestration}
\vspace{-0.25em}
Hyperscale cloud and AI operators routinely distribute computational workloads across multiple sites for redundancy, throughput, and latency optimization. Centralized orchestration frameworks (e.g., DeepSpeed, Kubernetes, or internal schedulers) synchronize training phases, checkpoint updates, and communication cycles across data centers connected to different transmission nodes \cite{kumar2025scalable, patel2024dynamic, sharma2024intelligent}.

When compute clusters at buses $i$ and $j$ initiate or pause tasks nearly simultaneously, their active power demand follows a phase-aligned temporal pattern, that can be approximated by:
\begin{align}
P_i(t) &= P_{i,0} + A_i \cos(\omega_c t + \phi_i), \\
P_j(t) &= P_{j,0} + A_j \cos(\omega_c t + \phi_j),
\end{align}
where $\omega_c$ is the characteristic frequency of the workload cycle. When $|\phi_i-\phi_j|\ll 1$, the cross correlation
\vspace{-0.75em}
\begin{equation}
\mathbb{E}[\Delta P_i(t)\Delta P_j(t)] = \frac{A_iA_j}{2}\cos(\phi_i-\phi_j)\approx\frac{A_iA_j}{2}
\end{equation}
is positive, which implies spatial correlation. This effect is particularly plausible in regions such as Northern Virginia, where a dense concentration of hyperscale data centers shares common transmission corridors with large commercial and residential loads. In such clusters, synchronized AI training schedules can create temporally correlated power fluctuations observable across multiple substations.

\begin{samepage}
\enlargethispage{1\baselineskip}
\vspace{-0.25em}
\subsection{Grid-Mediated Electromagnetic Coupling}
Modern AI data centers are \emph{converter-dominated loads} that interface with the grid through high-capacity AC/DC rectifiers, UPS systems, and DC--DC converters~\cite{sun2021development}. These converters exhibit fast voltage--current dynamics, low effective inertia, and nonlinear impedance characteristics~\cite{xiong2020modeling}. As a result, local power disturbances can travel through the network as electromagnetic effects that couple nearby substations~\cite{cui2022disturbance}.
\end{samepage}

Under small-signal conditions, the relation between nodal voltage and power perturbations can be approximated by
\vspace{-0.5em}
\begin{equation}
\Delta \mathbf{V} = \mathbf{Z}\,\Delta \mathbf{I}, \qquad 
\Delta I_k \approx \frac{\Delta P_k}{V_k^{*}}.
\vspace{-0.5em}
\end{equation}

The voltage deviation at bus~$j$ due to active-power variations across all buses is \vspace{-1em}
\begin{equation}
\Delta V_j = \sum_{k=1}^{N} Z_{jk}\frac{\Delta P_k}{V_k^{*}}.
\vspace{-0.5em}
\end{equation}
\needspace{6\baselineskip}
When a local power perturbation $\Delta P_i$ occurs at bus~$i$, converters at adjacent buses respond through grid-mediated coupling. The induced active-power response at bus~$j$ is
\begin{equation}
\Delta P_j^{\text{resp}} = -\frac{E'V\cos\theta}{X} \cdot \frac{\Delta P_i}{S_b},
\end{equation}
where $E'$ and $V$ are the internal and terminal voltages, $\theta$ is the power angle, $X$ is the line reactance, and $S_b$ is the system base power~\cite{cui2022disturbance}. The disturbance travels through the network as an \emph{electromechanical wave} with approximate speed
\begin{equation}
c \approx \sqrt{\frac{E' V \cos\theta\,\omega_s}{2 H S_b X}},
\end{equation}
where H is the system inertia. This links the fast voltage and current dynamics of spatially separated converter stations~\cite{cui2022disturbance}.

In regions with high data-center density, such as AI clusters sharing 230--500~kV corridors, this propagation mechanism synchronizes converter responses across substations. The resultant voltage–current alignment reinforces correlated load oscillations generated by synchronized computational workloads and reduces voltage stability margins.

\enlargethispage{1\baselineskip}
\subsection{Thermal and Environmental Coupling}
Beyond fast electrical interactions, slower environmental processes also generate correlated load patterns. Data centers located in the same geographical area experience similar weather conditions. Thus, their HVAC systems follow comparable heating and cooling patterns. The thermal dynamics of each facility is defined by a lumped-capacitance model~\cite{erden2013experimental}.
\begin{equation}
\vspace{-0.4em}
C_{\text{th}}\frac{dT_i}{dt}=P_{\text{IT}}^{(i)}(t)-\frac{T_i-T_{\text{amb}}}{R_{\text{th}}}-Q_{\text{cool}}^{(i)}(t),
\end{equation}
where $Q_{\text{cool}}^{(i)}$ is the HVAC cooling rate. 
When data centers experience similar ambient temperatures $T_{\text{amb}}$, their cooling controllers respond in a comparable manner. This leads to correlated cooling demand and electrical load patterns. 
Such effect is most pronounced in geographically clustered facilities but can also arise in large commercial or institutional buildings with similar HVAC control strategies.

The mechanisms above act on distinct timescales. GPU training induces high-frequency power fluctuations \cite{go2025characterizing}, converter interactions cause mid-frequency dynamics \cite{xiong2020modeling}, and HVAC systems add low-frequency modulation \cite{go2025characterizing}. When data centers are electrically and geographically proximate, these mechanisms may combine to form correlated load patterns.

\section{Impacts of Correlated Load on Power Systems}
\label{sec:Impacts}
Spatial and temporal correlations among large electrical loads reshape system dynamics and reliability margins. When load variations align across locations, aggregate disturbances rise, and system modes show stronger oscillations that reduce voltage stability, frequency balance, and available reserves.

\subsection{Voltage Stability Margin Degradation}
Voltage variability near an operating point is governed by the linearized power–flow sensitivities:
\begin{equation}
\begin{bmatrix}
\Delta P \\[2pt] \Delta Q
\end{bmatrix}
=
\begin{bmatrix}
J_{PP} & J_{PV}\\[2pt]
J_{QP} & J_{QV}
\end{bmatrix}
\begin{bmatrix}
\Delta \theta \\[2pt] \Delta V
\end{bmatrix},
\quad
\Delta V \approx -\,J_{QV}^{-1}\,\Delta Q,
\label{eq:lin_pf}
\end{equation}
so the smallest singular value $\sigma_{\min}(J_{QV})$ quantifies the weakest voltage–restoring stiffness.
The coupling between active- and reactive-power variations can be expressed through a local small-signal transfer function \(H_{QP,i}(j\omega)\) that captures converter admittance and network voltage sensitivity after the elimination of voltage perturbations \cite{sun2022dataI,sun2022dataII}.
At each bus \(i\),
\begin{equation}
\vspace{-.75em}
\Delta Q_i(j\omega) = H_{QP,i}(j\omega)\,\Delta P_i(j\omega). \label{eq: delta_Q}
\end{equation}
Expanding \(H_{QP,i}\) around the steady-state set point at (\(\omega\!=\!0\)):
\begin{equation}
H_{QP,i}(j\omega) = H_{QP,i}(0) + j\omega\,H'_{QP,i}(0) + \mathcal{O}(\omega^2). \label{eq: H_q eqn}
\end{equation}
Neglecting higher-order terms and defining
\begin{equation}
\alpha_i = H_{QP,i}(0), \qquad \beta_i = H'_{QP,i}(0), \label{eq: defining}
\end{equation}
the first-order approximation from \eqref{eq: delta_Q}, \eqref{eq: H_q eqn} and \eqref{eq: defining} becomes
\begin{equation}
\Delta Q_i(j\omega) \approx (\alpha_i + j\omega\beta_i)\,\Delta P_i(j\omega).
\end{equation}
Transforming to the time domain yields
\begin{equation}
\Delta Q_i(t) \approx \alpha_i\,\Delta P_i(t) + \beta_i\,\frac{d\Delta P_i(t)}{dt}.
\end{equation}
The coefficient \(\alpha_i\) represents the quasi-static dependence of reactive power on active power at the converter node, determined primarily by the DC-link and PLL control gains, while \(\beta_i\) quantifies the dynamic phase shift introduced by control and network time constants \cite{sun2022dataI,sun2022dataII,cui2022disturbance}.
\enlargethispage{6 pt}

Let the active-power fluctuations be spectrally correlated over a dominant frequency band $\Omega_c$, with correlation strength $\lambda_1$ and total spectral power $S_c$ at frequency $\omega_c$.
The corresponding voltage spectrum (for the most sensitive mode) can be approximated as
\begin{equation}
|\Delta V(j\omega)| \approx
\frac{|\Delta Q(j\omega)|}{\sigma_{\min}(J_{QV})} \approx
\frac{|\alpha + j\omega_c\beta|\, S_c\, \lambda_1}{\sigma_{\min}(J_{QV})}.
\label{eq:voltage_spectrum_approx}
\end{equation}
Voltage instability occurs when the correlated disturbance energy in the most sensitive mode equals the voltage-restoring capability, i.e.,$|\Delta V(j\omega_c)| \gtrsim 1.\label{eq:voltage_instability_condition}$
Rearranging \eqref{eq:voltage_spectrum_approx} yields the critical correlation threshold,
\begin{equation}
\lambda_1^{\text{crit}} =
\frac{\sigma_{\min}(J_{QV})}{
\sqrt{\alpha^2 + \omega_c^2 \beta^2}\, S_c}.
\label{eq:lambda_crit}
\end{equation}
Voltage instability is expected when $\lambda_1 > \lambda_1^{\text{crit}}$.
Equation~\eqref{eq:lambda_crit} expresses an energy balance between correlated stochastic excitation and voltage-restoring stiffness.
Higher load correlation ($\lambda_1$), dominant low-frequency content ($\omega_c$), strong active–reactive coupling ($\alpha,\beta$), or greater spectral power ($S_c$) all lower $\lambda_1^{\text{crit}}$, thereby reducing the voltage stability margin.

\subsection{Frequency Deviation Amplification}
Correlated load ramps reduce the effectiveness of inertial buffering~\cite{kundur2007power}.
When data centers impose synchronized active-power changes, the aggregate disturbance amplifies the initial frequency deviation.

Let $L_i$ denote zero-mean load fluctuations at bus $i$ with variance $\sigma^2$. For analytical tractability, the $N$ load fluctuations $L_i$ are assumed to have identical variances $\mathrm{Var}(L_i)=\sigma^2$.
The uniform pairwise correlation is
\begin{equation}
\rho_0 = \frac{\Cov(L_i,L_j)}{\sigma^2},
\vspace{-1em}
\end{equation}
so $\Cov(L_i,L_j) = \sigma^2$ if $i=j$, and $\rho_0\sigma^2$ otherwise.
The aggregate load $L_{\text{agg}} = \sum L_i$ then has variance
\begin{equation}
\begin{aligned}[t]
\sigma_{\text{agg}}^2
&= \mathrm{Var}\!\left(\sum_{i=1}^N L_i\right)
 = \sum_{i=1}^N \mathrm{Var}(L_i)
 + 2 \sum_{i<j} \mathrm{Cov}(L_i, L_j) \\[2pt]
\end{aligned}\notag
\end{equation}
\begin{equation}
\begin{aligned}
        &=\! N\sigma^2 + N(N\!-\!1)\rho_0\sigma^2
\! = \! N\sigma^2\!\left[1 + (N\!-\!1)\rho_0\right]
\end{aligned}
\label{eq:agg_var_corr}
\end{equation}

For a step $\Delta P_{\text{agg}} = \sqrt{1+(N-1)\rho_0}\,\Delta P_0$, the swing equation gives
\begin{equation}
\Delta f_0 \approx -\frac{\Delta P_0}{2H_{\text{sys}} f_0} \sqrt{1+(N-1)\rho_0}.
\label{eq:delta_f}
\end{equation}
where $f_0$ is the nominal system frequency and $H_{sys}$ is the system inertia constant. For $N=5$, $\rho_0=0.8$, the factor $\Delta f_0 \approx 2.05$, doubling the expected dip.
This reduces frequency margins under low-inertia conditions, where fast converter control contains electromechanical waves~\cite{cui2022disturbance} but synchronized action worsens RoCoF and nadir.

\vspace{-0.5em}
\subsection{Inter-Area Oscillation Amplification}
Spatially correlated load variations can couple into electromechanical oscillations of the interconnected system. 
When the spatial correlation mode of load fluctuations, say $\mathbf{u}_1$, aligns with a poorly damped inter-area mode shape $\boldsymbol{\phi}_k$, the modal excitation strength is proportional to the participation factor $|\mathbf{u}_1^\top \boldsymbol{\phi}_k|^2$. 
High values indicate constructive alignment between the correlated disturbance pattern and the oscillatory mode. 
Correlated activity from large-scale data centers or synchronized computing clusters, often concentrated around $\sim$1~Hz, may interact with inter-area modes typically occurring in the $0.2$–$0.8$~Hz range, potentially amplifying tie-line oscillations and degrading damping~\cite{kundur2007power,ko2025wide}. Spectral overlap can induce beat phenomena linking data center power electronics and AI workloads to wide-area oscillatory behavior \cite{mishra2025understanding, ko2025wide}. These effects are pronounced when the spectral content of the correlated load fluctuations overlaps the dominant electromechanical mode frequencies.

\vspace{-0.5em}
\subsection{Planning and Operating Reserve Implications}
Traditional diversity factors assume that individual load fluctuations are statistically independent, which yields an aggregate variance that scales linearly with the number of buses. Spatially distributed data center loads, on the other hand exhibit correlated behavior. 
Using the covariance structure derived in~\eqref{eq:agg_var_corr}, positive correlation ($\rho_0>0$) increases aggregate variability relative to the independent-load aggregate. 
For illustration, with $N{=}10$ and a moderate correlation level $\rho_0{=}0.3$ (30\% correlation), the aggregate variance is $37 \times$ larger than that of a single bus and $3.7\times$ larger relative to the independent-load baseline $N\sigma^2$, corresponding to a $\sqrt{3.7}\approx 1.9\times$ increase in aggregate standard deviation. This enlarged variability necessitates proportionally higher operating and contingency reserves to maintain equivalent reliability margins under correlated demand fluctuations~\cite{kundur2007power}.

\section{Real-Time Digital Simulation Studies}
\label{sec:RTDS}
Real-time simulations were conducted on the IEEE 39-bus test system using the RTDS platform to assess the dynamic impact of spatially correlated data center loads. RTDS was selected because it maintains real-time execution and synchronous phase relationships across loads (which offline tools cannot guarantee) and provides sub-microsecond timestep resolution to capture converter dynamics. Three representative AI data centers were placed at buses 4, 12, and 15, each supported by dedicated battery energy storage systems (BESS), with ratings summarized in Table~\ref{tab:AIloadBESS}. Dynamic load models with average-value converter representations were employed for the data centers following \cite{Chaudhary2025DataCenterStability}, rather than detailed switching models, as switching frequencies (5--20~kHz) are far above the correlation time scales of interest. The active and reactive power demands evolved according to multi-cycle AI workload patterns, with a correlation coefficient of $\rho \approx 0.4$ imposed across the three sites to model partial synchronization driven by shared computational scheduling. This moderate correlation represents a conservative middle ground between statistical independence ($\rho$ = 0) and perfect synchronization ($\rho$ = 1). 

\vspace{-0.5em}
\begin{table}[htbp]
\centering
\caption{AI data center loads and BESS rating}
\label{tab:AIloadBESS}
\resizebox{\columnwidth}{!}{
\begin{tabular}{lcccc}
\hline
Bus & $P$ [MW] & $Q$ [MVAr] & $P_{\text{BESS}}$ [MW] & $E_{\text{BESS}}$ [MWh] \\
\hline
4  & 200 & 40.6 & 200 & 50.0 \\
12 & 225 & 45.7 & 225 & 56.3 \\
15 & 250 & 50.8 & 250 & 62.5 \\
\hline
\end{tabular}
}
\end{table}

\begin{figure*}[!htbp]
    \centering
    \subfloat[Active and reactive power variations\label{fig:load_profile}]{
        \includegraphics[width=0.31\textwidth]{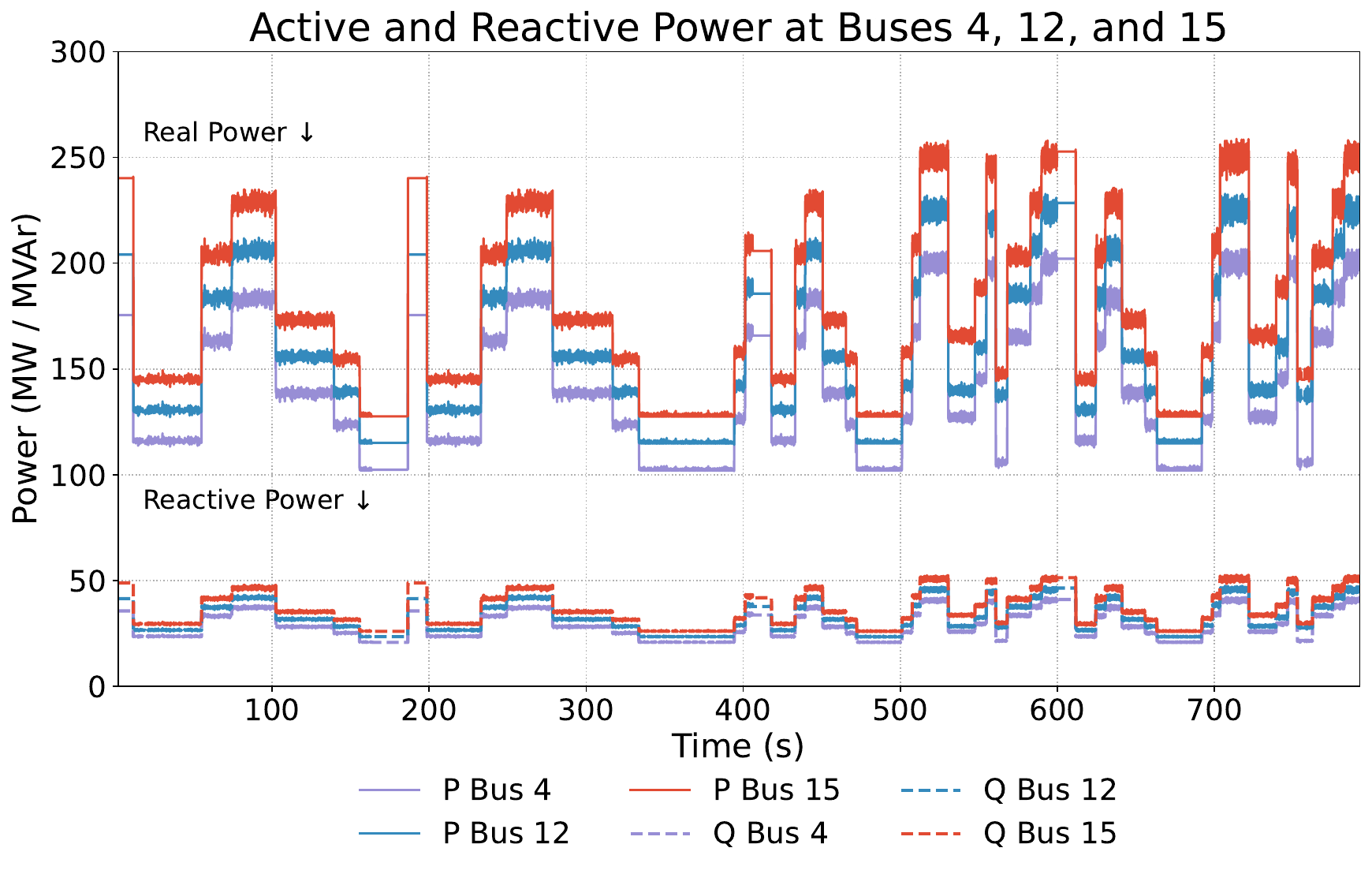}
    }
    \hfill
    \subfloat[System frequency response\label{fig:freq_response}]{
        \includegraphics[width=0.31\textwidth]{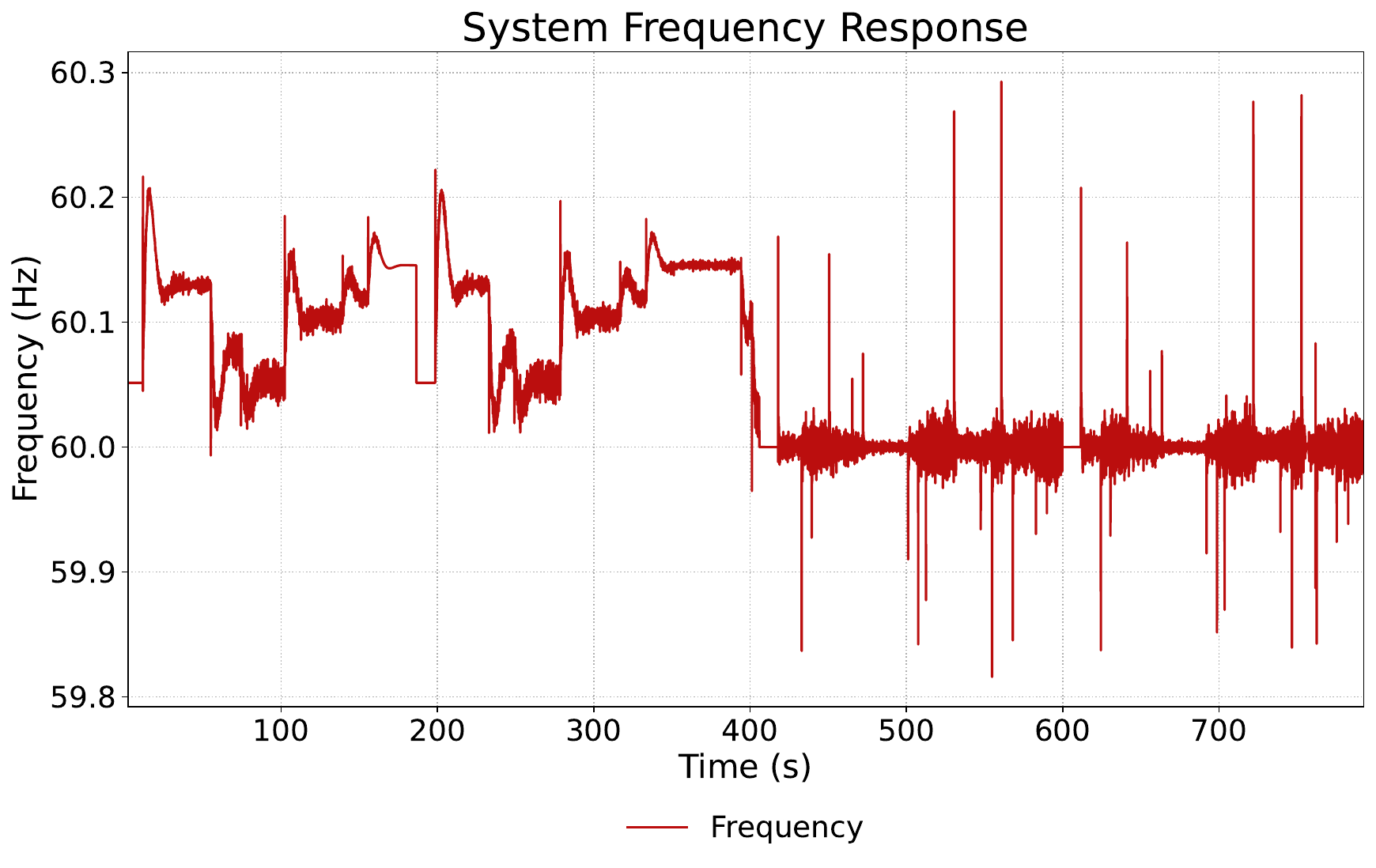}
    }
    \hfill
    \subfloat[RMS voltage fluctuations\label{fig:voltage_rms}]{
        \includegraphics[width=0.31\textwidth]{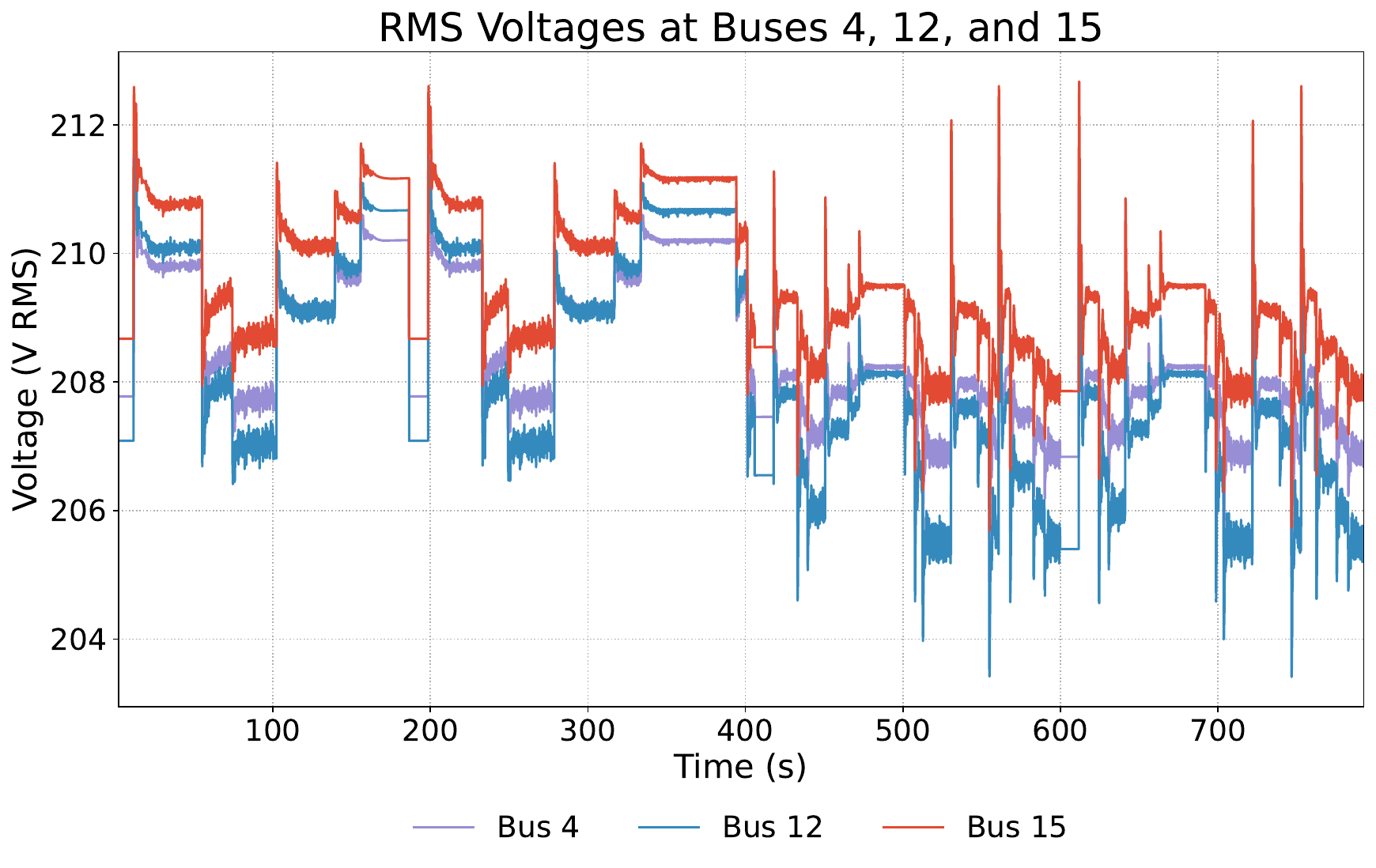}
    }
    \caption{Dynamic responses of the IEEE 39-bus system under correlated data-center load profiles}
    \label{fig:system_responses}
    \vspace{-1em}
\end{figure*}

Figure~\ref{fig:load_profile} shows the active and reactive power trajectories at the data center buses. The traces reveal synchronized ramp-up and ramp-down sequences, with active power peaks occurring nearly concurrently across all sites. Reactive power exhibits analogously but with minor phase shifts from independent converter voltage regulation. This temporal coherence in load evolution generates amplified aggregate power swings relative to uncorrelated scenarios, consistent with theoretical expectations. The system frequency response, shown in Fig.~\ref{fig:freq_response}, displays clear sensitivity to the correlated load ramps. During coincident demand surges, frequency deviates downward by up to 0.15~Hz from the 60~Hz nominal. Recovery occurs promptly as load decreases, supported by generator droop action and BESS discharge. The distributed BESS units limit the total frequency excursion to within 0.3 Hz peak-to-peak. This further signifies the need for dedicated ESS for effective suppression of frequency deviations induced by correlated load fluctuations.
Figure~\ref{fig:voltage_rms} illustrates the RMS voltage response at buses~4, 12, and~15. Voltage magnitudes remain within $\pm2\%$ of the nominal value but show distinct fluctuations during the correlated load ramps. The nearly simultaneous voltage dips observed across all three buses reveal a shared electromechanical mode driven by spatially correlated load variations. This behavior demonstrates inter-bus coupling in the voltage domain and underscores the influence of coherent load fluctuations on network-wide voltage stability.

Real-time simulations confirm that moderate load correlation ($\rho \approx 0.4$) markedly diminishes the natural diversity effect that underpins classical load independence assumptions. Incorporating spatial load correlation into transient analysis becomes essential to capture realistic grid behavior. Coordinated BESS operation proves vital to maintain frequency and voltage stability in data-center-dominated networks.

\section{Conclusion}
\label{sec:Conclusion}
This paper provides a scientific rationale for characterizing and analyzing spatially correlated loads in data-center-dominated power systems. Analytical derivations quantify how load correlation amplifies aggregate disturbances, weakens voltage stability margins, degrades frequency response, and excites inter-area oscillation modes. Real-time digital simulations on the IEEE 39-bus system validate theoretical predictions and demonstrate that moderate spatial correlation produces simultaneous frequency and voltage deviations across multiple buses. The results confirm that traditional diversity assumptions no longer hold in converter-dominated grids with synchronized computational workloads.

Future research will focus on empirical validation with data from data-center-intensive regions to calibrate correlation models and refine stability thresholds. The development of real-time load correlation estimation algorithms for system operators represents a operational need. Investigation of coordinated control strategies for distributed energy storage systems to mitigate correlation-induced oscillations offers an important direction for grid management. Finally, integration of load correlation metrics into long-term transmission planning tools will support infrastructure investment decisions in the evolution of AI-driven power systems.

\balance
\section*{Acknowledgment}
The research is supported in part by Sandia National Lab, in part by MSU Research Foundation and in part by the U.S. National Science Foundation under grant 2408615.
\vspace{-1em}
\bibliography{references}
\bibliographystyle{ieeetr}

\end{document}